\documentclass[useAMS,usenatbib]{mn2e}
\usepackage{aas_macros}
\usepackage[dvips]{graphicx}
\usepackage{amsmath,amssymb}
\title[The effects of baryon physics, black holes and AGN feedback in clusters of galaxies]{The effects of baryon physics, black holes and AGN feedback on the mass distribution in clusters of galaxies}
\author[D. Martizzi et al.]{\parbox[t]{\textwidth}{Davide Martizzi$^{1}$\thanks{E-mail: martdav@physik.uzh.ch}, 
Romain Teyssier$^{1,2}$, Ben Moore$^{1}$ and Tina Wentz$^{1}$}\vspace*{6pt}\\
$^{1}$Institute for Theoretical Physics, University of Zurich, CH-8057 Z\"urich, Switzerland\\
$^{2}$CEA Saclay, DSM/IRFU/SAP, B\^atiment 709, F-91191 Gif-sur-Yvette, Cedex, France}

\begin{document}

\maketitle

\label{firstpage}

\begin{abstract}
The spatial distribution of matter in clusters of galaxies is mainly determined by the dominant dark matter component, however, physical processes involving baryonic matter are able to modify it significantly. We analyse a set of 500 pc resolution cosmological simulations of a cluster of galaxies with mass comparable to Virgo, performed with the AMR code RAMSES. We compare the mass density profiles of the dark, stellar and gaseous matter components of the cluster that result from different assumptions for the subgrid baryonic physics and galaxy formation processes. First, the prediction of a gravity only N-body simulation is compared to that of a hydrodynamical simulation with standard galaxy formation recipes, then all results are compared to a hydrodynamical simulation which includes thermal AGN feedback from Super Massive Black Holes (SMBH). We find the usual effects of overcooling and adiabatic contraction in the run with standard galaxy formation physics, but very different results are found when implementing SMBHs and AGN feedback. Star formation is strongly quenched, producing lower stellar densities throughout the cluster, and much less cold gas is available for star formation at low redshifts. At redshift $z=0$ we find a flat density core of radius 10 kpc in both of the dark and stellar matter density  profiles. We speculate on the possible formation mechanisms able to produce such cores and we conclude that they can be produced through the coupling of different processes: (I) dynamical friction from the decay of black hole orbits during galaxy mergers; (II) AGN driven gas outflows producing fluctuations of the gravitational potential causing the removal of collisionless matter from the central region of the cluster; (III) adiabatic expansion in response to the slow expulsion of gas from the central region of the cluster during the quiescent mode of AGN activity. 
\end{abstract}

\begin{keywords}
black hole physics -- cosmology: theory -- cosmology: large-scale structure of Universe -- galaxies: formation -- galaxies: clusters: general -- methods: numerical
\end{keywords}

\section{Introduction}

Clusters of galaxies are the most massive virialised structures observed in the Universe and provide a wonderful laboratory to test astrophysical theories. In the $\Lambda$CDM cosmological scenario, clusters are assembled via a hierarchy of mergers of less massive structures like galaxies and groups of galaxies. Many physical processes play a role during the formation of a cluster. When satellite galaxies are accreted into a cluster, their properties can be changed by tidal and ram pressure stripping, leading to the formation of a wide variety of galaxy morphologies. Furthermore, clusters are known to be dark matter dominated structures, with most of the baryonic matter residing in a hot diffuse X-ray emitting gaseous phase, the Intracluster Medium (ICM). The stellar mass is less significant and mainly contained in the massive central elliptical galaxy. 

Since they are dominated by dark matter, this mass component determines the global properties of the mass distribution in the cluster. However, from the theoretical side, it is well known that baryonic processes can produce significant differences in the distribution of matter in collapsed structures with respect to models including only collisionless cold dark matter. For example, baryons are known to condense the centre of dark matter halos due to dissipative processes, producing adiabatic contraction of the total mass distribution  \citep{Gnedin:2004p569}. Several models including baryonic physics have been invoked to solve the so--called cusp/core problem in the $\Lambda$CDM cosmological framework, i.e. the discrepancy between the centrally cuspy dark matter profiles observed in dark matter halos in numerical N-body simulations and the centrally cored dark matter profiles inferred by observations in dwarf galaxies and low surface brightness galaxies (see the recent review by \citealp{2010AdAst2010E...5D} and references therein). The study of baryon physics induced modifications in the mass distribution in collapsed structures, and clusters of galaxies in particular, is still a field with many open issues. 

The present paper is dedicated to the study of the effects of baryonic processes on the mass distribution in clusters of galaxies. In particular, we use a set of cosmological hydrodynamical simulations performed using the AMR code RAMSES \citep{Teyssier:2002p451} to study the effect of different models for baryons and galaxy formation physics on the mass density profile of a cluster of galaxies comparable to the Virgo cluster. This work can be considered as an extension of the analysis performed by \cite{2011MNRAS.414..195T} on the same simulations, and is complementary to the analysis presented by \cite{2012MNRAS.420.2859M}. Here, we focus on the peculiar properties produced in the mass density profile when including Super Massive Black Holes (SMBHs) and the related AGN feedback in the recipes for galaxy formation physics. We stress that AGN feedback was initially introduced to solve the so--called "overcooling problem", namely the fact that too much stellar mass is produced in massive collapsed structure in hydrodynamical simulations with respect to what is observed in the real Universe \citep{Borgani:2009p728}. The strong quenching of star formation produced by processes that couple AGN activity with the gas is expected to improve the match between simulations and observations \citep{Tabor:1993p1080, Ciotti:1997p1087, Silk:1998p941}. Strong evidence for the existence of AGN feedback is provided by observations of X-ray cavities and radio blobs in galaxy clusters, typically interpreted as buoyantly rising bubbles of high entropy material injected in the central region of clusters by jets of relativistic particles. In this paper, we show that by including SMBH physics and AGN feedback it is possible to obtain interesting predictions on the modifications they can induce on the mass distribution in massive dark matter halos.

The paper is organized as follows: the first section is dedicated to the numerical methods and the galaxy formation recipes adopted for our simulations; we show our main results and provide our interpretation in the second section; the last section is left for a short summary of our results and to a discussion.

\section{Numerical methods}
\label{sec:num_methods}

In this section, we describe the numerical techniques used to model the cluster we consider in this paper. We consider two cosmological hydrodynamical simulations presented in \cite{2011MNRAS.414..195T} and \cite{2012MNRAS.420.2859M}, plus a third cosmological simulation with only dark matter. For all the three runs we used the AMR code RAMSES \citep{Teyssier:2002p451}. The simulations were performed using the zoom-in technique which allows to obtain the required effective resolution only in selected regions of the computational domain. For all the three simulations the computational domain is a cubic box of side 100 Mpc/h. For the dark matter only run (dubbed as DMO, Table \ref{tab:cosm_par} and \ref{tab:mass_par}) we adopt a standard $\Lambda$CDM cosmology with parameters $\Omega_{\rm m}=0.3$, $\Omega_\Lambda=0.7$, $\Omega_{\rm b}=0.0$, $\sigma_{\rm 8}=0.77$ and $H_0=70$ km/s/Mpc. For the two hydrodynamical runs (HYDRO runs in Table \ref{tab:cosm_par} and \ref{tab:mass_par}) we adopt the same values for $\Omega_{\rm m}$, $\Omega_\Lambda$, $\sigma_{\rm 8}$ and $H_0$, but we set  $\Omega_{\rm b}=0.045$. The initial conditions for the three simulations where computed using the \cite{Eisenstein:1998p1104} transfer function and the {\ttfamily grafic} package \citep{Bertschinger:2001p1123} in its parallel implementation {\ttfamily mpgrafic} \citep{Prunet:2008p388}. To perform the zoom-in technique we adopted the following approach: first, we ran a low resolution dark matter only simulation, then we identified dark matter halos at $z=0$. From this halo catalog we built a set of halos whose virial masses lie in the range $10^{14}$ to $2 \times 10^{14}$ M$_\odot$/h. Finally, we identified the final halo based on its assembly history: the halo is already in place at $z=1$, therefore it can be considered as relaxed at $z=0$. {In particular the last major merger is observed at redshift $z\sim 1.3$.} The final virial mass is $M_{\rm vir} \simeq 10^{14}$~M$_\odot$, while $M_{\rm 200c}=1.04 \times 10^{14}$~M$_\odot$ or $M_{\rm 500c}=7.80 \times 10^{13}$~M$_\odot$, where indice $c$ refers to the critical density. This halo has been re-simulated at higher resolution, focusing the computational resources in the region of the computational domain where it forms.

\begin{table}
\begin{center}
{\bfseries Cosmological parameters}
\begin{tabular}{|l|c|c|c|c|c|}
\hline
\hline
 {\itshape Type} & $H_0$ [km s$^{-1}$Mpc$^{-1}$] & $\sigma_{\rm 8}$ & $\Omega_\Lambda$ & $\Omega_{\rm m}$ & $\Omega_{\rm b}$ \\
\hline
\hline
 DMO & 70 & 0.77 & 0.7 & 0.3 & - \\
 HYDRO & 70 & 0.77 & 0.7 & 0.3 & 0.045 \\
\hline
\hline
\end{tabular}
\end{center}
\caption{Cosmological parameters adopted in our simulations. The DMO label refers to the dark matter only run. The HYDRO label refers to the hydrodynamical runs.}\label{tab:cosm_par}
\end{table}

In the DMO simulations the dark matter particle in the high resolution region has a mass of $9.6 \times 10^6$ M$_\odot$/h. In the two hydrodynamical runs, the dark matter particle in the high resolution region has a mass of $8.2 \times 10^6$ M$_\odot$/h, whereas the baryon resolution element has a mass of $1.4 \times 10^6$ M$_\odot$/h. Hydrodynamics is solved on an AMR grid that was initially refined to the same level of refinement than the particle grid ($2048^3$, level $\ell=11$). During the runs, 7 more levels of refinement were considered (level $\ell_{\rm max}=18$). The refinement criterion we used allows spatial resolution to be roughly constant in physical units; the minimum cell physical size was always close to $\Delta x_{\rm min} = L/2^{\ell_{\rm max}}\simeq 500$ pc/h. The grid was dynamically refined using a quasi-Lagrangian strategy: when the dark matter or baryons mass in a cell reaches 8 times the initial mass resolution, it is split into 8 children cells.

\begin{table}
\begin{center}
{\bfseries Mass and spatial resolution}
\begin{tabular}{|l|c|c|c|c|}
\hline
\hline
{\itshape Type} & $m_{\rm cdm}$&  $m_{\rm gas}$ & $m_*$ & $\Delta x_{\rm min}$ \\
 & $[10^{6}$ M$_\odot$/h] & $[10^{6}$ M$_\odot$/h] & $[10^{6}$ M$_\odot$/h] & [kpc/h] \\
\hline
\hline
 DMO & $9.6$ & n.a. & n.a. & $0.38$ \\
 HYDRO & $8.2$ & $1.4$ & $0.3$ & $0.38$ \\
\hline
\hline
\end{tabular}
\end{center}
\caption{Mass resolution for dark matter particles, gas cells and star particles, and spatial resolution (in physical units) for our 2 sets of simulations. }\label{tab:mass_par}
\end{table}

In the HYDRO runs, gas dynamics is modeled using a second-order unsplit Godunov scheme \citep{Teyssier:2002p451, Teyssier:2006p413, Fromang:2006p400} based on the HLLC Riemann solver and the
MinMod slope limiter \citep{Toro:1994p1151}. We assume a perfect gas equation of state (EOS) with $\gamma=5/3$. The HYDRO runs include standard subgrid models for gas cooling (accounting for H, He and metals; we use the \citealt{sutherland_dopita93} cooling function), star formation (we adopt a star formation efficiency $\epsilon_*=0.01$) and supernovae feedback (we adopt the "delayed cooling" scheme, \citealt{Stinson:2006p1402}) and metal enrichment. In one of the two hydrodynamical simulations we also implement AGN feedback, using a modified version of the \cite{Booth:2009p501} model. SMBHs are modeled as sink particles, following the prescriptions of \cite{Krumholz:2004p1079}, and they are allowed to merge. Gas accretion onto each SMBHs is computed adopting a modified Bondi-Hoyle formula \citep{Booth:2009p501}. A fraction of the accreted mass is converted into thermal energy that is directly injected into the gas surrounding the black hole. For practical purposes, in the rest of the paper we will refer to the run with AGN feedback as AGN-ON, and to the run without AGN feedback as AGN-OFF. Further details about the galaxy formation and AGN feedback recipes can be found in \cite{2012MNRAS.420.2859M} and \cite{2011MNRAS.414..195T}. 

{As discussed in \cite{2011MNRAS.414..195T}, the AGN feedback energy is effectively deposited in two different modes: the energetic and impulsive ‘quasar mode’ (Eddington limited luminosity $\sim 5 \times 10^{46}$ erg s$^{-1}$) during cold gas accretion when the accretion rate is high, and the quiescent ‘radio mode’ during hot gas accretion when the accretion rate is low (Bondi-Hoyle-limited luminosity $\sim 5 \times 10^{41}$ erg s$^{-1}$). At $z\sim 0$ the central SMBH in the AGN-ON cluster accretes mass in radio mode at a rate $\sim 10^{-4}$ M$_{\odot}$ yr$^{-1}$, producing a total luminosity of $9 \times 10^{40}$ erg s$^{-1}$, comparable to the X-ray 
luminosity of the AGN in M87 $L_{X,0.5-7 keV} \approx 7 \times 10^{40}$ erg s$^{-1}$ \citep{Allen:2006p5821}; the mechanical power of the jet of the AGN in M87 is much larger, $P \approx 3.3 \times 10^{43}$ erg s$^{-1}$. These considerations imply that the AGN in M87 lies in an intermediate state between the radio and the quasar modes of our model. Further indications that the activity of the central SMBH is not particularly atypical are provided by comparison of the accretion rates and star formation rates in the AGN-ON simulation with observational results at redshift $z\sim 0$. The mass accretion rate onto the SMBH is $\sim 10^{-4}$ M$_{\odot}$ yr$^{-1}$ and the star formation rate within the inner few kpc from the cluster center is $\sim 10^{-2}$ M$_{\odot}$ yr$^{-1}$. These values are comparable with those recently measured by \cite{2012ApJ...746..168D} for Seyfert 1 galaxies.}

\section{Results}
\label{sec:results}

In this Section we show how SMBHs and AGN feedback are able to change dramatically the properties of the mass distribution in clusters of galaxies. We focus on 3D mass density profiles of the cluster, analysing separately the dark matter and stellar components, as well as the total mass distribution. The goal is to highlight differences between the profiles in the DMO, AGN-ON and AGN-OFF run. The interpretation of our main results are provided in the next Section. 

\subsection{Mass density profiles at redshift $z=0$}

Spherically averaged mass density profiles of the cluster at redshift $z=0$ have been computed for all the simulations. A reliable identification of halo centres is required to compute density profiles. Halo centres and virial radii have been identified using the AdaptaHOP algorithm \citep{2004MNRAS.352..376A}, in the version implemented and tested by \cite{2009A&A...506..647T}, using the Most Massive Substructure Method (MSM) to identify halos as well as their substructures. AdaptaHOP can be used to find the centres of groups of dark matter particles (i.e. ``halos'') as well as groups of star particles in the HYDRO runs (i.e. ``galaxies''). In the DMO run we directly use the centre of the dark matter halo, whereas in the AGN-ON and AGN-OFF runs we use the centre of the most massive group of star particles (i.e. the centre of the brightest cluster galaxy) since it traces better the minimum of the gravitational potential. To test that the latter choice does not lead to errors in the estimate of the profiles in the inner region of the cluster, we verified that picking the halo centre instead of the galaxy centre does not change the measured profiles.

\begin{figure*}
    \includegraphics[width=0.99\textwidth]{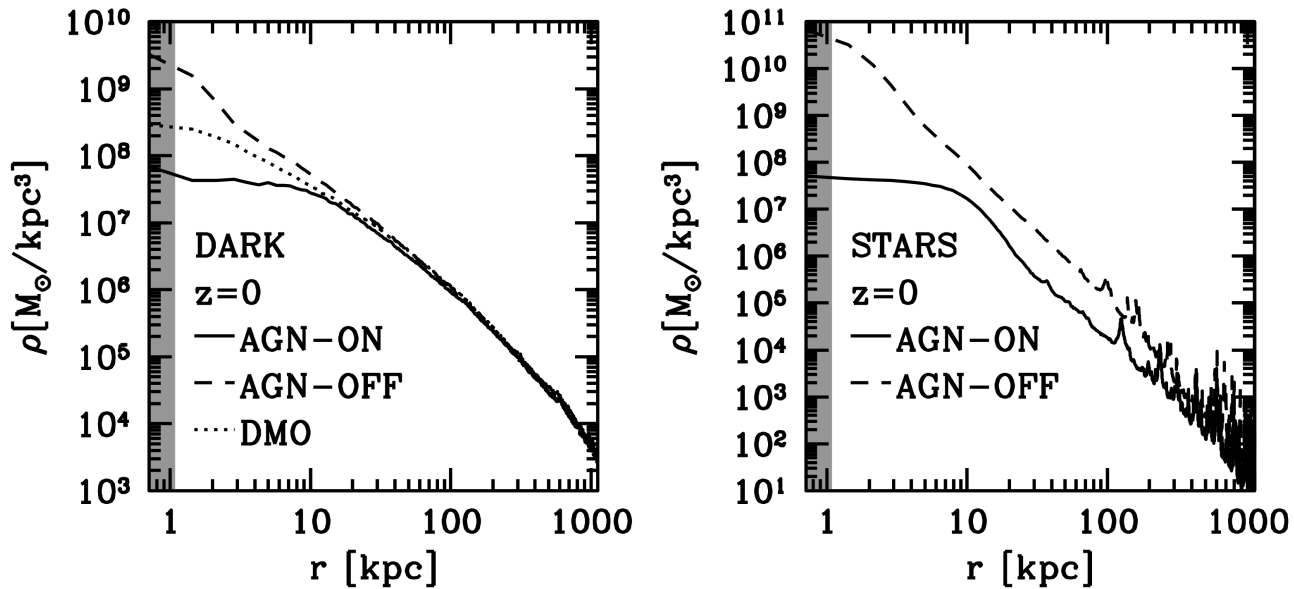}
  \caption{ Mass density profiles at $z=0$. Left: dark matter. Right: stars. In all panels the
  grey shaded area represents the spatial resolution.}
  \label{fig:profiles_z0}
\end{figure*}

Figure \ref{fig:profiles_z0} shows the dark matter and stellar mass density profiles at $z=0$ for the AGN-ON (solid line) and AGN-OFF (dashed line) runs; in this plot we use physical units. In the left panel the result of the DMO run (dotted line) is also plotted for comparison. All the profiles extend to the virial radius of the cluster. The three dark matter profiles look quite similar in the outer regions ($r\gtrsim 50$ kpc), while they are significantly different in the inner regions. The AGN-OFF profile is much more centrally peaked and concentrated than the DMO profile, a result that can be interpreted as a result of adiabatic contraction of dark matter halos in response to the condensation of baryons at their centres \citep{Gnedin:2004p569, 2011MNRAS.414..195T}; we test the quality of adiabatic contraction models in reproducing our results in the Subsection \ref{sec:fits}. The AGN-ON profile is much different, with a very shallow inner slope. Within the inner 10 kpc we observe a dark matter core whose size is much larger than our spatial resolution limit (grey shaded area), whereas the AGN-OFF and DMO profiles seem to be consistent with cuspy dark matter profiles. Observational support to the results of our AGN-ON model comes from studies where gravitational lensing and dynamical data are combined to estimate the dark matter profiles of clusters of galaxies  \citep{2004ApJ...604...88S, 2008ApJ...674..711S, 2009ApJ...706.1078N, 2011ApJ...728L..39N, 2011A&A...531A.119R}: these studies provide significative evidence of the existence of clusters of galaxies with dark matter profiles presenting cores or shallower inner slopes than in the Navarro-Frenk-White model. 

Looking at the stellar mass density profiles, we see that significantly less stellar mass forms in the AGN-ON run with respect to the AGN-OFF run that is affected by overcooling. At $r>50$ kpc the stellar mass density in the AGN-OFF run is $\sim 10$ times bigger than in the AGN-ON run in all the radial bins, and a very significative difference can be observed between the two profiles within $\sim 10$ kpc from the centre: the AGN-OFF density profile is increasing towards the centre, whereas the AGN-ON profile shows a flat core with a size comparable to that observed in the dark matter profile. It is interesting that cores in the stellar surface brightness profiles of very massive elliptical galaxies and cluster central galaxies have been observed by several authors \citep{1999ASPC..182..124K, 2000ApJS..128...85Q, 2003AJ....125..478L, 2004AJ....127.1917T, 2005AJ....129.2138L, 2007ApJ...671.1456C,  2009ApJS..182..216K, 2011arXiv1108.0997G} and they can be deprojected in cored 3D stellar density profiles \citep{2005MNRAS.362..197T}.

In the AGN-OFF cluster the mass budget in the central region is strongly dominated by stellar mass: in absence of AGN feedback the star formation process is very efficient in turning gas into stars; the large stellar mass in the central region of the AGN-OFF cluster has been assembled through cooling flows providing gaseous material to trigger intense star formation events and through the accretion of satellite galaxies onto the central galaxy. In contrast, the mass in the central regions of the AGN-ON cluster has comparable contributions from stars and dark matter: overcooling is prevented from happening by AGN feedback and star formation is strongly quenched.

\subsection{Dark matter profiles at $z=0$: testing the adiabatic contraction model}\label{sec:fits}

\begin{figure*}
    \includegraphics[width=0.99\textwidth]{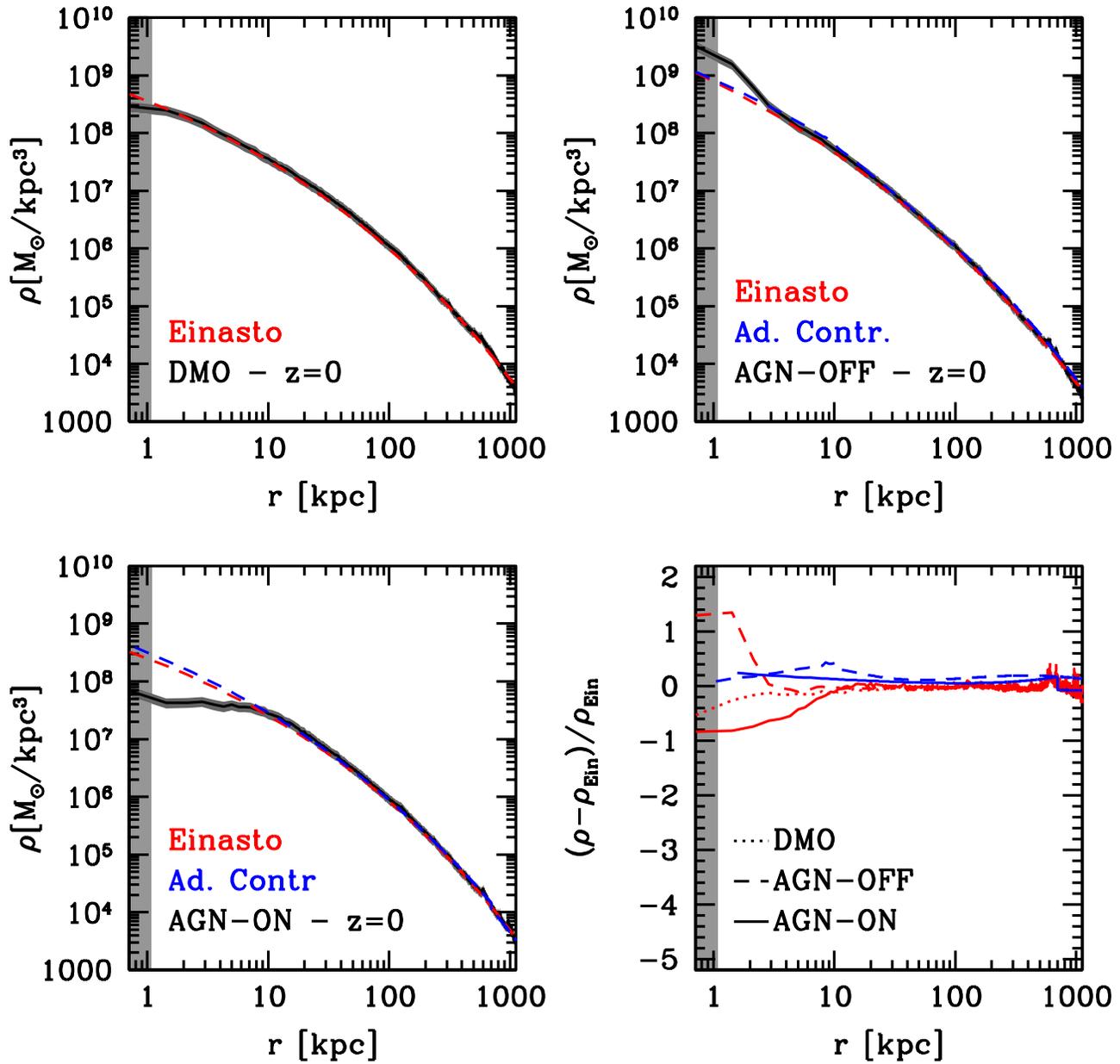}
 \caption{ Comparison between the dark matter profiles measured at redshift $z=0$ (black solid lines) and the Einasto fits (red dashed lines). The scatter due to Poisson noise in each radial bin is also represented as a dark grey shaded area. The DMO Einasto fit is adiabatically contracted to get the two blue dashed profiles. Top-left panel: DMO run. Top-right panel: AGN-OFF run. Bottom-left panel: AGN-ON run. Bottom-right panel: relative difference between the measured profile and the Einasto fit as a function of radius (red lines); we use a dotted line for the DMO run, a dashed line for the AGN-OFF run and a solid line for the AGN-ON run; the blue lines show the residuals of the adiabatic contraction models with respect to the Einasto fits. In all panels the grey shaded area represents the spatial resolution.}
  \label{fig:mass_excess}
\end{figure*}

\begin{center}
\begin{table*}
\begin{tabular}{|l|c|c|c|c|}
\multicolumn{5}{|c|}{{\bfseries Einasto Profile Fits - $z=0$}} \\
\hline
\hline
{\itshape Simulation} & $\rho_{\rm e} [M_{\odot}/kpc^3]$&  $r_{\rm e} [kpc]$ & $n$ & $\widetilde{\chi}^2$ \\
\hline
\hline
 DMO & $1.43\times10^3\pm 1.5\times10^2$ & $1.60\times10^3\pm 6\times10^1$ & $5.93\pm 1.2\times10^{-1}$ & 0.94 \\
 AGN-ON & $1.39\times10^3\pm 1.1\times10^2$ & $1.55\times10^3\pm 5\times10^1$ & $5.65\pm 9\times10^{-2}$ & 0.97 \\
 AGN-OFF & $6.2\times10^2\pm 9\times10^1$ & $2.0\times10^3\pm 1.1\times10^2$ & $7.38\pm 1.8\times10^{-1}$ & 2.21 \\
\hline
\hline
\end{tabular}
\caption{Best fit parameters $\rho_{\rm e}$, $r_{\rm e}$ and $n$ for the Einasto profile at redshift $z=0$. The value of the reduced chi-squared $\widetilde{\chi}^2$ (1386 d.o.f.) for $10\hbox{ kpc}<r<1000\hbox{ kpc}$ is also reported. The values are reported for the DMO, AGN-ON and AGN-OFF simulations.}
\label{tab:profiles}
\end{table*}
\end{center}
It is possible to estimate some of the effects of baryonic processes, SMBHs and AGN feedback using an approach similar to that used by observers when analysing the surface brightness profiles of galaxies. First, an analytical model is used to fit the profile, then any significant deviation from the model is interpreted as a signature of physical processes. The same approach has been used to detect central light excesses/deficiencies with respect to a S{\'e}rsic fit to the surface brightness profiles of early-type galaxies \citep{2009ApJS..182..216K, 2011arXiv1108.0997G}. In \cite{2012MNRAS.420.2859M} we also adopted this approach to discuss the properties of the stellar core observed in the stellar mass density profile in the AGN-ON run, showing that the S{\'e}rsic function can be used to fit the stellar mass surface density profile outside the cored region. Here, we use this criterion to analyse the dark matter profiles we measure in our simulations at $z=0$. 

We adopt the Einasto profile as our fiducial analytical model, since it has been shown to provide excellent fits to the dark matter profiles observed in cosmological N-body simulations \citep{2005ApJ...624L..85M, 2006AJ....132.2701G}. We use the following parameterization:
\begin{equation}
\rho_{\rm Ein}(r)=\rho_{\rm e}\exp\left\{-d_{\rm n}\left(\frac{r}{r_{\rm e}} \right)^{1/{\rm n}}-1\right\}
\end{equation}
where 
\begin{equation}
d_{\rm n}=3n-1/3-0.0079/n.
\end{equation}
We use this analytical function to fit the dark matter profiles at $z=0$, leaving $I_{\rm e}$, $n$ and $r_{\rm e}$ as free parameters. Our fits are performed using the Levenberg-Marquardt nonlinear least squares fit algorithm \citep{Press:1992p1847}. The parameters we obtain after the fits are summarized in Table \ref{tab:profiles}. {In the DMO case we measure a concentration $c_{\rm 200}=R_{\rm 200c}/R_{\rm -2}=5.88$, where $R_{\rm -2}$ is the radius at which the logarithmic slope of the dark matter profile is -2. This value is typical for halos of mass $\sim 10^{14}$ M$_{\odot}$ \citep{2011MNRAS.415.3177R}. }

The fits are compared to the measured profiles in Figure \ref{fig:mass_excess}. The Einasto profile provides a very good fit to the dark matter profile of our halo in the DMO run, despite the fact that it is typically used to fit the averge profile in cosmological simulations whereas we analyse only one halo. Both in the AGN-ON and AGN-OFF we observe that the Einasto profile is a good fit for $r>10$ kpc, while we observe significative deviations with respect to the fitting formula in the inner regions (see the bottom-right panel of Figure \ref{fig:mass_excess}). The presence of these features can be interpreted as a manifestation of processes that influence the formation of a standard distribution of dark matter in phase space; since these features are not observed in the gravity-only DMO run, we try to give them an explaination in terms of baryon physics. 

\begin{table}
\begin{center}
{\bfseries Adiabatic Contraction Model Parameters }
\begin{tabular}{|l|c|c|}
\hline
\hline
{\itshape Simulation} & $m_{\rm d}$ [$M_{\odot}$] & $r_{\rm d}$ [kpc]  \\
\hline
\hline
 AGN-ON & 1.7$\times 10^{13}$ & 700 \\
 AGN-OFF & 2.6$\times 10^{11}$ & 10 \\
\hline
\hline
\end{tabular}
\caption{Parameters adopted for the adiabatic contraction model: truncated sphere mass $m_d$ and radius $r_d$ . }\label{tab:ac}
\end{center}
\end{table}

Dark matter halos are expected to respond adiabatically to the condensation of baryons at their centres \citep{Gnedin:2004p569, 2011MNRAS.414..195T}. We compare the results of our hydrodynamical runs at $z=0$ with the prediction of a simple adiabatic contraction model used by \cite{Abadi:2009p531}, and already adopted in \citep{2011MNRAS.414..195T}; details for this model can be found in Appendix \ref{appendix:A}. We adiabatically contract the Einasto fit to the DMO profile assuming that the baryons are distributed in a constant surface density, truncated sphere of mass $m_d$ and radius $r_d$, despite the fact that the actual distribution is different. The values of $m_d$ and $r_d$ adopted in this paper are shown in Table \ref{tab:ac}. The result is plotted in Figure \ref{fig:mass_excess} (blue lines): in both the AGN-ON and AGN-OFF cases the adiabatically contracted profiles match the Einasto fits quite well at all radii, but, like the fits, the model does not provide a satisfactory description of the measured profiles for $r<10$ kpc. A mass excess with respect to the adiabatic contracted profile is detected in the AGN-OFF case. 
The adiabatic contraction model works reasonably well, especially at radii beyond a few kpc. {In the interval $10\hbox{ kpc}<r<1000\hbox{ kpc}$ we measure reduced chi-squared values for the adiabatic contraction models $\widetilde{\chi}^2=2.14$ for the AGN-OFF case and $\widetilde{\chi}^2=0.72$ for the AGN-ON case. These values are quite similar to what we obtain for the Einasto fits in the same radial range (Table \ref{tab:profiles})}.


In the AGN-ON run we find a mass deficiency with respect to the adiabatically contracted profile within 10 kpc from the centre; this mass deficiency is $M_{\rm def}=5.7\times10^{10}$ M$_{\odot}$. The fact that adiabatic contraction works for the outer regions of the cluster, but cannot predict the formation of the core, highlights the fact that additional processes play a role in shaping the properties of the mass distribution. Given the theoretical and computational limits of our phenomenological model for AGN feedback and the formation of SMBHs, and given that we limit our analysis to the evolution of one halo, these results provide evidence that SMBH physics is indeed relevant in shaping the properties of clusters of galaxies, especially in the most overdense regions where SMBHs are expected to be found. We will discuss these topics in further detail in subection \ref{sec:form_core}. 

\subsection{Evolution of the mass density profiles}\label{sec:evol_profs}

The study of the evolution of the mass density profiles in our simulations can be used to shed more light on the mechanisms through which SMBHs and AGN feedback influence the mass distribution in clusters. To do so, we found the centres of the most massive progenitors of the cluster in each simulation, computed density profiles at redshifts $z=$ 1, 2, 3, 4, 5 and then compared to the result at $z=0$. All the profiles shown in this subsection are plotted in physical units.

\begin{figure}
    \includegraphics[width=0.49\textwidth]{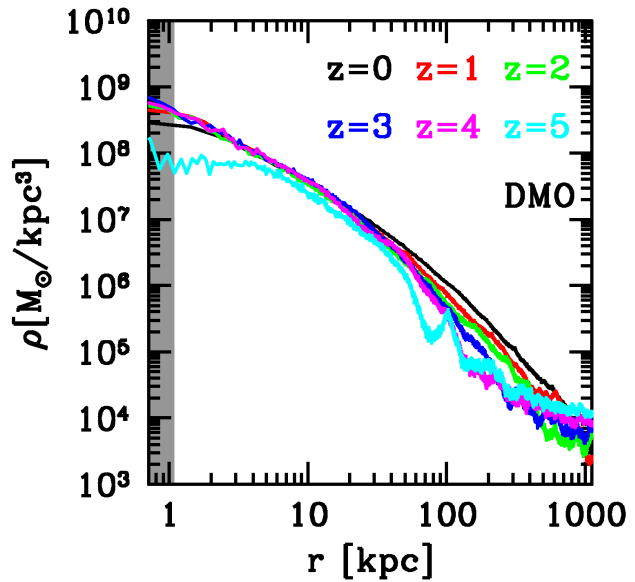}
 \caption{ Evolution of the dark matter density profile in the DMO run from $z=5$ to $z=0$. The
  grey shaded area represents the spatial resolution.}
  \label{fig:prof_dmo}
\end{figure}

First of all, we analyse the dark matter profile evolution in the DMO simulation (Figure \ref{fig:prof_dmo}).  In the outskirts of the cluster, the profiles change their slope; this transition marks the virial radius of the cluster. At redshift $z=5$ the mass distribution in the halo is still quite different than at later times, with a shallower central slope and a radius $R_{\rm 200c}\approx 60$ kpc. At later times the inner part of the profiles ($r\lesssim 20$ kpc) reaches stability and evolution can be observed only in the outskirts where additional mass collapses into the halo, in fact the high $r$ tail of the profile is constantly extending towards bigger distances from the centre. At redshift $z=0$ we measure that $R_{\rm 200c}\approx 1$ Mpc. This kind of evolution is a clear exemple of stable clustering. 

\begin{figure*}
    \includegraphics[width=0.99\textwidth]{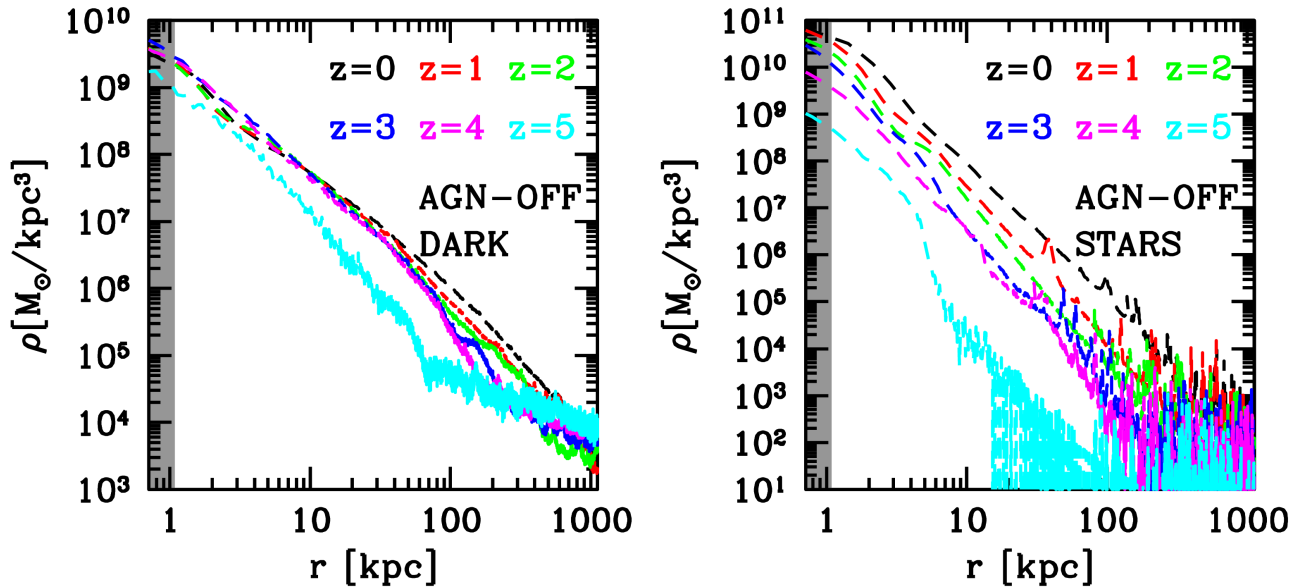}
 \caption{ Evolution of the density profiles in the AGN-OFF run from $z=5$ to $z=0$. Left: dark matter. Right: stars. In all panels the
  grey shaded area represents the spatial resolution.}
  \label{fig:prof_noagn}
\end{figure*}

In Figure \ref{fig:prof_noagn} we show the evolution of the dark and stellar mass density profiles in the AGN-OFF run. The effect of adiabatic contraction due to the condensation of baryonic mass at the centre of the cluster is already evident at redshift $z=5$: the density within the inner 10 kpc is almost an order of magnitude larger than in the DMO case. Similarly to the DMO case, as the dark matter halo is assembled, the dark matter profile in the inner region appears to be stable from $z=4$ to $z=0$, while additional dark mass appears to be accreted at higher radii. The stellar mass profiles evolves mantaining its shape, but increasing in amplitude. New stars are constantly formed and a significant increase of the stellar mass can be observed between all the considered snapshots. 

\begin{figure*}
    \includegraphics[width=0.99\textwidth]{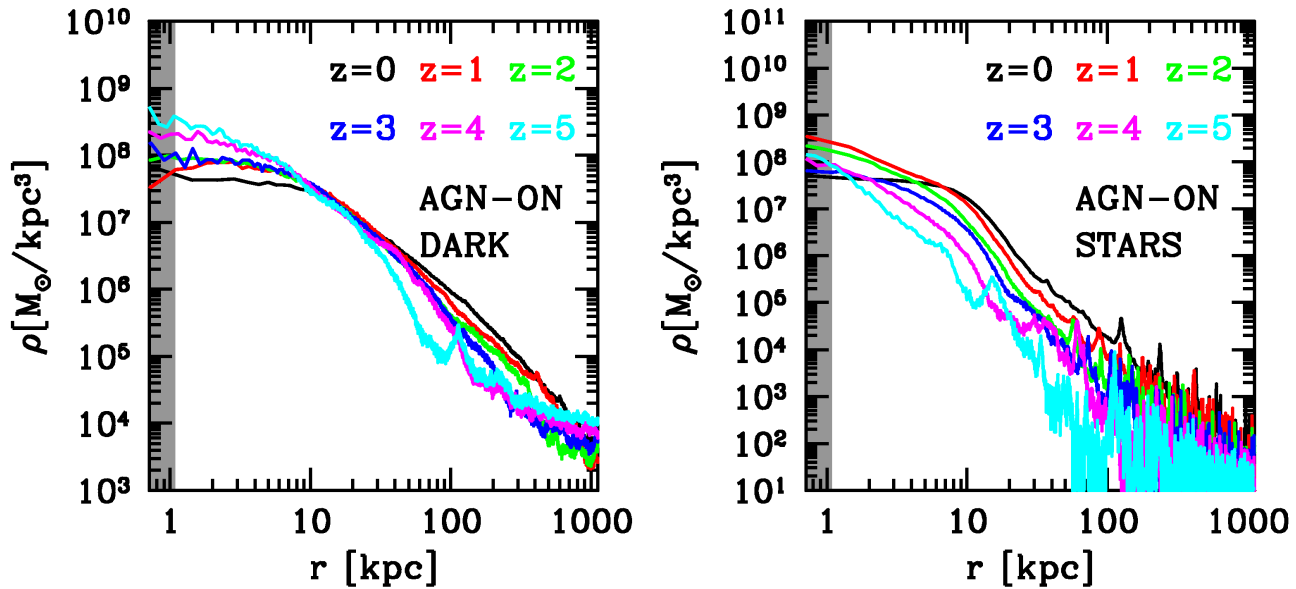}
 \caption{ Evolution of the density profiles in the AGN-ON run from $z=5$ to $z=0$. Left: dark matter. Right: stars. In all panels the grey shaded area represents the spatial resolution.}
  \label{fig:prof_agn}
\end{figure*}

Finally, Figure \ref{fig:prof_agn} shows the same profiles for the AGN-ON run, highlighting some of the most interesting peculiarities of our model. The evolution of the dark matter profile in the outer regions of the cluster ($r\gtrsim 20$ kpc) is very similar to that observed in the DMO case, but the inner region is strikingly different. The effects of the processes leading to the formation to the core observed at $z=0$ can be directly observed. At redshift $z=5$ the dark matter profile appears to be contracted with respect to the DMO case, but much less than in the AGN-OFF case. Instead of mantaining the original slope, like in the AGN-OFF case, the inner part of the dark matter profiles becomes shallower with time, until a core is formed. A very similar behaviour is observed for the stellar mass density profile. While stellar mass is always increasing with time at radii $r\gtrsim 20$, the inner profile evolves in a very peculiar way: at redshift $z=5$ the stellar profile is very cuspy, but it evolves becoming shallower and shallower, despite the total stellar mass within $10$ kpc from the centre increases. Between $z=1$ and $z=0$ the stellar profile becomes almost completely flat for $r<10$ kpc. 

\subsection{ Formation of central cores in the density profile }\label{sec:form_core}

The most evident peculiarity of the mass distribution in our AGN-ON simulation is the central core observed in the stellar and dark matter density profiles. From the plots (Figure \ref{fig:prof_agn}) in the previous subsection we see that the dark matter core forms more gradually than the stellar core. At $z>1$ we see that the dark matter profile gradually evolves towards a cored configuration, while the stellar profile builds up mantaining its cuspy inner shape. The most interesting transition towards the formation of the two cores happens between redshifts $z=1$ and $z=0$: in this interval the dark matter profile becomes extremely flat, while the cusp at the centre of the stellar profile is completely erased. The fact that this behaviour is not observed in the AGN-OFF simulation suggests that the cores in the AGN-ON dark and stellar matter profiles do not form naturally, but they are generated through external processes influencing the dynamics of collisionless matter. The aim of this subsection is to elucidate the mechanisms that lead to core formation in our simulated cluster.

Several models have been proposed to produce cores in the density profile of a distribution of collisionless matter starting from a cuspy profile; these models were originally developed to explain the dark matter cores observed in gas rich dwarf galaxies \citep{2010AdAst2010E...5D}, but may be applied, in principle, to any distribution of collisionless matter. \cite{2001ApJ...560..636E} showed that sufficiently massive gas clumps can disrupt cusps through dynamical friction; this is not observed in our AGN-ON run since gas clouds are easily disrupted by AGN feedback. \cite{2004ApJ...607L..75E} showed that galaxies moving within the dark matter background and transferring their orbital energy to the dark matter via dynamical friction may contribute to the formation of cores. More recently, the very high resolution simulation of \cite{Naab:2009p5731} showed a similar effect: repeated minor dry mergers lead to decreases of the central stellar density concentration. {This result has been confirmed by subsequent numerical experiments, like those presented by \cite{2012arXiv1202.2357L}}. The dense cores of accreted galaxies able to survive for a few crossing times modify the stellar mass distribution through dynamical friction. We emphasise that the same principle applies to the dark matter component, due to its collisionless nature. This effect is not observed in our AGN-OFF run because completely dry mergers are rare due to the high gas fractions in galaxies in this simulation. In the AGN-ON run, instead, a significant fraction of the gas is expelled from galaxies and the \cite{Naab:2009p5731} mechanism can be more efficient. Unfortunately, it seems to be challenging to produce an extended flat core like the one observed in our AGN-ON model only through repeated dry minor mergers. 

Alternative mechanisms to produce cores in dark matter profiles involve purely gravitational processes related to SMBHs. In the context of a $\Lambda$CDM cosmology, where massive structures form through the hierarchical mergers of less massive structures, the formation of SMBH binaries is an expected result. At the centre of collapsed objects SMBHs form binary pairs whose orbits decay as they transfer their orbital energy to collisionless matter via three-body interactions: the result is that collisionless matter can be expelled from the central regions via the gravitational slingshot effect and a core is formed. This process is usually referred to as SMBH scouring \citep{2003ApJ...596..860M} and it is expected to remove $\sim 2-4$ times the mass of the SMBH formed after the binary completely decays \citep{2007ApJ...671...53M}. SMBH scouring is important on spatial scales from 100 to 1 pc, thus it cannot be resolved in our simulations which has 1 kpc force softening. At the softening length the mass is completely dominated by dark matter and stars so that SMBH binaries cannot form. 

\begin{figure}
    \includegraphics[width=0.49\textwidth]{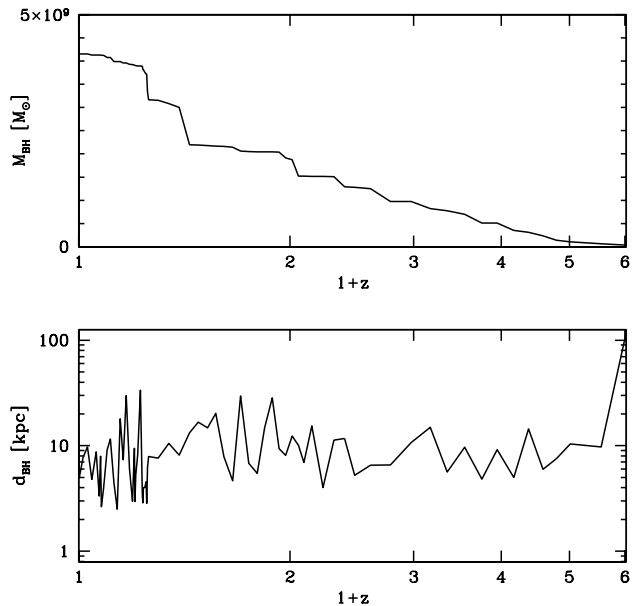}
 \caption{ Top: Evolution of the central SMBH observed at redshift $z=0$. Bottom: Evolution of the distance between the central SMBH observed at redshift $z=0$ and its closest neighbour. }
  \label{fig: black hole}
\end{figure}

Despite the fact that SMBH scouring is not resolved in our simulations, they are able to resolve another process able to produce cores, namely SMBHs sinking to the very central region due to dynamical friction during mergers. The efficiency of this process has been extensively studied by \cite{2010ApJ...725.1707G} and their results show that the orbital energy tranferred from the SMBHs to collisionless matter contributes to the formation of cores. The efficiency of this process is not as high as in the SMBH scouring, but it is able to produce mass deficiencies comparable to the mass of the infalling SMBHs. In our AGN-ON simulation the mass of the central SMBH at $z=0$ is $M_{\rm BH}=4.2\times 10^9$ M$_{\odot}$, so  this process by itself is not able to explain the mass deficiencies we observe in the central regions of the cluster: $M^{\rm dark}_{\rm def}=5.7\times10^{10}$ M$_{\odot}$ for dark matter and $M^{\rm star}_{\rm def} = 3.04\times10^{10}$ M$_{\odot}$ for stars, measured with respect to an Einasto fit and a S{\'e}rsic fit to the entire profile, respectively. 

Recently, the pure N-body simulations performed by  \cite{2011arXiv1107.0517K} showed that when multiple (two or more) SMBHs are present in a halo, the core formation process is much more efficient than during the inspiral of a single SMBH. Considering this enhanced core formation via dynamical friction it is possible to create mass deficits of more than 5 times the total SMBH mass. The top panel of Figure \ref{fig: black hole} shows the mass growth of the central SMBH at the centre of the cluster at $z=0$. The mass growth of this black hole is very smooth for $z>2$, but it proceeds mainly through major mergers (sudden jumps in the black hole mass) at lower redshifts. The bottom panel of Figure \ref{fig: black hole} shows the distance between the central SMBH and its closest neighbour as a function of redshift, dubbed as $d_{\rm BH}$. Any sudden increase in the SMBH mass $M_{\rm BH}$ due to a black hole merger in the top panel plot can be detected as an increase in $d_{\rm BH}$, because after mergers the second neighbour becomes the first neighbour. The continuous variation of $d_{\rm BH}$ tells us that at any redshift black holes are moving in the halo and losing orbital energy because of dynamical friction. A significative number of major mergers of very massive black holes happens at $z<1$. This fact suggests that the \cite{2011arXiv1107.0517K} mechanism is particularly efficient at redshifts $z<1$.

SMBH infall is not the only process contributing to the formation of the core. In the AGN-ON simulation strong AGN driven outflows are observed \citep{2012MNRAS.420.2859M}. AGN feedback greatly increases the local temperature and entropy of gas, then the high entropy material is tranported out of the central region of the cluster through convective motions. These outflows modify the local gravitational potential and may cause expansion of both the dark and stellar mass distribution. Similar processes have been observed in numerical simulations in which gas outflows generated by supernovae feedback are used to produce cores in the dark matter profiles of dwarf galaxies \citep{1996MNRAS.283L..72N, 2002MNRAS.333..299G, 2005MNRAS.356..107R, 2010Natur.463..203G, 2011arXiv1106.0499P}. In the simulations performed by \cite{1996MNRAS.283L..72N}, the mass outflows are simulated by growing and rapidly removing an idealised potential from the centre of an equilibrium realisation of a dark matter halo, showing that the natural consequence is the formation of a core. The efficiency of this core formation process is $\propto M_{\rm disc}^{1/2}R_{\rm disc}^{-1/2}$, where $M_{\rm disc}$ is the mass of the disc and $R_{\rm disc}$ is its scale radius. 

\begin{figure}
    \includegraphics[width=0.49\textwidth]{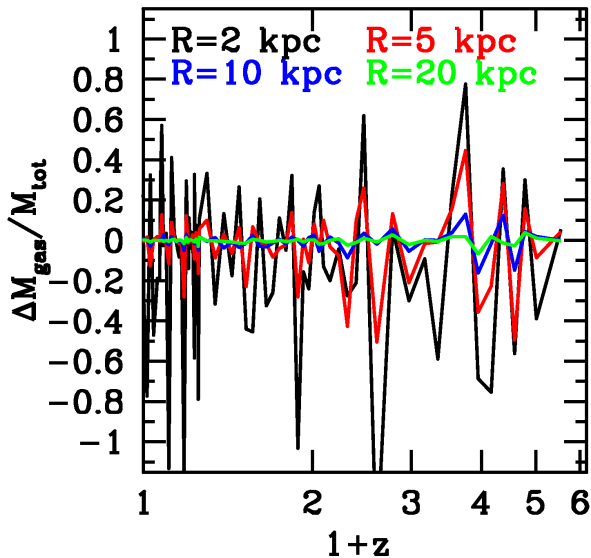}
 \caption{ Variation of the gas mass enclosed in spheres of radius $R=2, 5, 10 \hbox{ and } 20$ physical kpc with respect to the total mass in the same region. The absolute value of the fluctuations can be larger than 1 because the variation of the gas mass is divided by the total mass after the outflow.}
  \label{fig:gmass_fluct}
\end{figure}

The calculations performed by \cite{2002MNRAS.333..299G} show that this mechanism is extremely inefficient when a single supernova explosion is considered, however \cite{2005MNRAS.356..107R} showed that repeated explosions followed by gas outflows and subsequent infalls are able to account for the formation of dark matter cores. The recent numerical experiments performed by \cite{2010Natur.463..203G} and \cite{2011arXiv1106.0499P} confirm the efficiency of this mechanism in a fully cosmological context. In particular, \cite{2011arXiv1106.0499P} suggest that gravitational potential fluctuations induced by supernovae driven outflows happening on a timescale shorter than the dynamical time cause the expansion of the dark matter distribution and the formation of a core. For this mechanism to be effective it is required to have gas outflows able to remove a significant fraction of the mass enclosed in the region where the core forms. Since we have multiple epochs of AGN driven gas outflows, this mechanism is likely to be active also in our AGN-ON run. This is indeed the case: Figure \ref{fig:gmass_fluct} shows the gas mass fluctuations induced by AGN burst driven outflows can be high fractions of the total mass in the central regions. Regions close to the central SMBH (radius $r<R=2-5$ kpc) present extreme mass fluctuations on short timescales. Within $r<R=10$ kpc fluctuations become smaller, but can be as large as 10\% of the enclosed total mass. At higher distances from the centre the effect of these mass fluctuations is almost undetected. This means that potential fluctuations will be particularly strong only within the inner 10 kpc from the centre, that is the region where the core is observed. It is interesting to note that the amplitude of the mass fluctuations is typically higher at $z>1$, however quite strong fluctuations are also observed at low redshift.

\begin{figure}
    \includegraphics[width=0.49\textwidth]{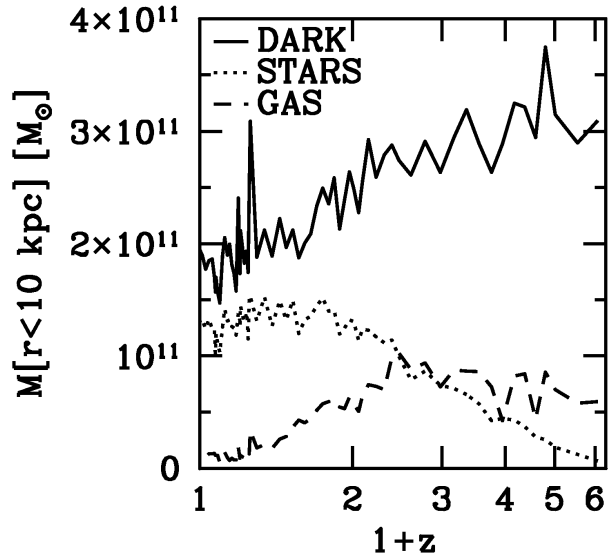}
 \caption{ Time evolution of the total dark, stellar and gas mass enclosed within 10 physical kpc in the AGN-ON run. }
  \label{fig:mass_evo}
\end{figure}

\begin{figure}
    \includegraphics[width=0.49\textwidth]{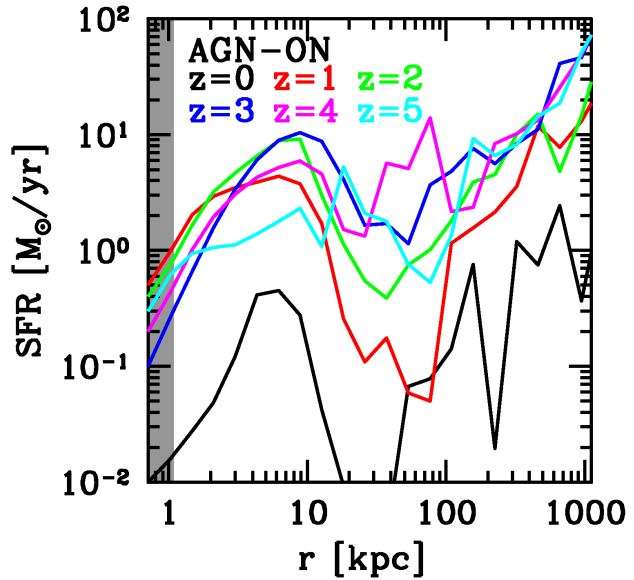}
 \caption{ Average star formation rate in radial bins in the AGN-ON run. The
  grey shaded area represents the spatial resolution.}
  \label{fig:sfr_agn}
\end{figure}

Before drawing our conclusions, we carefully analyse what happens in the central region of the cluster. Figure \ref{fig:mass_evo} shows how the mass distribution in the central region of the cluster is influenced by the core formation mechanisms. We plot the evolution of the mass enclosed within 10 physical kpc for all the different components: dark matter (solid lines), stars (dotted lines) and gas (dashed lines). The core formation processes produce a decrease in the dark mass within 10 kpc from the centre from $\sim 3\times 10^{11}$ M$_{\odot}$ at $z=5$ to $\sim 2\times 10^{11}$ M$_{\odot}$ at $z=0$. The stellar mass in the centre increases with time only down to $z=1$, staying approximately constant until $z=0$. At high redshift the star formation rate in the central region is high and the concentration of the stellar mass distribution is boosted by star formation events (Figure \ref{fig:sfr_agn}); at redshift $z<1$ the star formation in the central galaxy is strongly quenched by AGN feedback (Figure \ref{fig:sfr_agn}), so the concentration of the stellar distribution cannot be boosted by strong star formation events, thus letting the core formation processes be effective. At $z<1$ when AGN feedback becomes very strong because of the presence of very massive black holes, a slow decrease in the gas mass in the centre is observed: a gas mass $\sim 10^{11}$ M$_{\odot}$ is slowly removed from the central region before $z=0$. This slow decrease in gas mass is expected to produce an adiabatic expansion of the total mass distribution, which will also contribute to the formation of a central core. Furthermore, the cooling time of hot gas within the inner 30 kpc of the cluster centre is $\sim 1$ Gyr, suggesting that in its quiet mode the AGN can slowly eject the gas that rains down onto the centre from the inner cooling flow. 

The general picture we find from our analysis seems to show that the interplay between dynamical processes connected to SMBHs and the effects of AGN feedback on the star formation and the gas spatial distribution may provide an explaination to what is observed in our simulations. The dark matter core starts forming at redshift $z=4-5$ due to AGN burst driven gas mass fluctuations on short timescales. The process goes on with very high efficiency down to redshift $z=1$. At redshift $z < 1$ the infall of very massive SMBHs contributing to the core formation process compensates for the smaller amplitude of the gas mass fluctuations. The final result is a flat dark matter core at redshift $z=0$. The formation of the stellar core proceeds quite differently, with star formation compensating for the ejection of stellar material at $z>1$, preventing the early formation of a core. At $z<1$ the decrease in the star formation rate caused by AGN feedback finally allows the stellar core to form by redshift $z=0$.

Given our conclusions, we must stress that a clearer picture on the core formation problem has still to be drawn. The problem needs to be studied within a larger class of halos with different merger histories and masses, and idealised simulations are probably required to shed more light on the core formation process in clusters. 

\section{Summary and Conclusions}
\label{sec:summary}

We used the results of a set of cosmological simulations to study the effect of baryons in massive dark matter halos, focusing on phenomena involving SMBHs and AGN feedback. We use the zoom-in technique to simulate the formation of a galaxy cluster of mass comparable to Virgo at high spatial and mass resolution. In this work we analyse three different simulations: the first is a gravity only run with no baryons (DMO), the second is a hydrodynamical run with standard galaxy formation physics (AGN-OFF), the third is a hydrodynamical run with standard galaxy formation physics plus SMBHs and AGN feedeback (AGN-ON). We adopt a modified version of the \cite{Booth:2009p501} model to implement thermal AGN feedback in the AGN-ON run. This model allows to reproduce the self-regulated growth of black holes throughout the cosmic ages and it is tuned to reproduce the observed $M_{\rm BH}-\sigma$ relation. 

Our analysis focuses on the evolution of the 3D mass density profiles of the cluster from redshift $z=5$ to redshift $z=0$. We find a number of interesting differences between our models that highlight the importance of accounting for all the physical processes taking place during the formation and evolution of the most massive bound structures in the Universe. Our results can be summarized in the following points: 

\begin{itemize}
 \item At redshift $z=0$ the dark matter density profiles of the three simulations are consistent with each other at distances $r>10$ kpc from the centre, but they differ significantly in the central region. The DMO profile is consistent with a standard Einasto model. The AGN-OFF profile shows a mass excess with respect to an Einasto fit due to adiabatic contraction of the baryons at the centre of the cluster. The AGN-ON profile presents a flat core of radius 10 kpc, i.e. a mass deficiency with respect to an Einasto fit. The dark matter core forms gradually from $z=5$ to $z=0$. We stress that dark matter cores in dark matter profiles have been claimed to be observed in several clusters \citep{2004ApJ...604...88S, 2008ApJ...674..711S, 2009ApJ...706.1078N, 2011ApJ...728L..39N, 2011A&A...531A.119R}.

\item The stellar density profile at $z=0$ is also very different between the AGN-OFF and AGN-ON runs. The AGN-OFF profile is very peaked at the centre, while the AGN-ON profile has a core of the same size as that of the dark one. Unlike the dark core, the stellar one forms between $z=1$ and $z=0$. Note that, cores in the surface brightness profiles of massive ellipticals and cluster central galaxies have been observed by several authors \citep{1999ASPC..182..124K, 2000ApJS..128...85Q, 2003AJ....125..478L, 2004AJ....127.1917T, 2005AJ....129.2138L, 2007ApJ...671.1456C,  2009ApJS..182..216K, 2011arXiv1108.0997G}, so our model is not in disagreement with observations and it {\it predicts} the existence of dark matter cores associated to stellar cores in massive ellipticals. Furthermore, in the AGN-OFF run the stellar density is larger than in the AGN-ON run at all radii. This is a clear effect of overcooling of gas leading to enhanced (and too efficient) star formation. In the AGN-ON run, AGN feedback strongly quenches star formation and the overcooling problem is avoided  
\citep{2011MNRAS.414..195T,2012MNRAS.420.2859M}. The result is also that less cool gas is available at the centre of the AGN-ON cluster with respect to the AGN-OFF run: gas is heated by AGN feedback and carried away via convective motions and slow adiabatic expansion. 

 \item The core in the dark matter end stellar mass density profiles is a peculiar feature of our simulation that includes the physics associated with active galactic nuclei. In this paper, we suggest that the coupling of several mechanisms is responsible for their formation. First, SMBHs transfer part of their orbital energy to collisionless matter via dynamical friction during dry galaxy mergers \citep{2010ApJ...725.1707G}, especially at redshift $z<1$. Second, AGN driven gas outflows modify the gravitational potential in regions close to SMBHs with resulting ejection of collisionless matter from the central region of the cluster; subsequent gas outflows followed by the central 'revirialisation' of the central material are expected to produce cores \citep{2011arXiv1106.0499P}; due to the stronger AGN feedback at $z>1$, this mechanism is more effective at high redshift, but preserves an important role at low redshift. Third, the central hot gas slowly cools radiatively, falling onto the SMBHs in convective flows and is subsequently ejected impulsively; the slow loss of mass from the central region will produce an expansion of the inner mass distribution. 
\end{itemize}

Despite assertions that AGN feedback can affect central cluster dynamics, in fact
our results seem to be complementary to the those obtained by other authors for less massive dark matter halos \citep{Governato:2010p1442, 2011arXiv1106.0499P, 2011arXiv1111.5620M}.  Related processes involving baryons seem to be active at the low and high mass end of the galaxy mass function, especially mechanisms involving gas outflows produced by feedback. 

Given the fact that we perform the analysis on zoom simulations of one cluster, we need to stress that our result needs support from a suite of dedicated cosmological simulation aimed at exploring the effect of SMBHs, AGN feedback and baryon physics in general on the most massive clusters in the universe. Furthermore, a number of numerical experiments is also needed to explore in detail the core formation processes that have been considered in our discussion, with particular attention on their coupling. {The role of baryon mass outflows in galaxy clusters has been recently studied by \cite{2012arXiv1202.1527R} using collisionless matter simulations where gas is modeled as a time varying external contribution to the gravitational potential; their results provide important support to ours.} To conclude, we are convinced that our results are robust 
enough to assert that the implementation of AGN feedback and SMBHs in cosmological hydrodynamical simulations is an important ingredient for modeling massive clusters of galaxies.

\section*{Acknowledgments}
The AMR simulations presented here were performed on the Cray XT-5 cluster at CSCS, Manno, Switzerland.


\bibliography{papers}


\appendix
\section{Adiabatic Contraction Model}\label{appendix:A}

The simplified model we adopt is based on that used in \cite{2011MNRAS.414..195T}. If one defines the initial radius of each dark matter shell as $r_{\rm i}$, then an adiabatic contraction model is able to predict its value after contraction $r_{\rm f}$. In our case we adopt the transformation
\begin{equation}\label{eq:AC}
 \frac{r_{\rm f}}{r_{\rm i}}=1+\alpha\left(\frac{M_{\rm i}}{M_{\rm f}}-1\right)
\end{equation}
where $M_{\rm i}$ and $M_{\rm f}$ are the cumulative mass distributions before and after contraction, respectively. The final cumulative mass distribution can be computed as
\begin{equation}\label{eq:AC1}
 M_{\rm f}=M_{\rm dm}(r_{\rm f})+M_{\rm bar}(r_{\rm f})=f_{\rm dm}M_{\rm i}(r_{\rm i})+M_{\rm bar}(r_{\rm f}).
\end{equation}
where $M_{\rm i}(r_{\rm i})$ is the initial mass distribution in the DMO case, $M_{\rm bar}(r_{\rm f})$ is the baryonic mass distribution and $M_{\rm dm}(r_{\rm f})$ is the adiabatically contracted dark matter distribution. The dark mass fraction is computed as $f_{\rm dm}=1-m_{\rm d}/M_{\rm 200}$. Our aim is to recover the contracted dark matter profile $M_{\rm dm}(r_{\rm f})$ given $M_{\rm i}(r_{\rm i})$ and $M_{\rm bar}(r_{\rm f})$. We assume that the $M_{\rm i}(r_{\rm i})$ can be described by the fit to the Einasto profile we obtain for the DMO run at redshift $z=0$:
\begin{equation}
 M_{\rm i}(r_{\rm i})=4\pi\int_0^{r_{\rm i}}r^2\rho_{\rm Ein}(r)dr.
\end{equation}
The baryonic mass distribution is modeled as a constant surface density sphere with size $r_{\rm d}$ and mass $m_{\rm d}$:
\begin{equation*}
M_{\rm bar}(r_{\rm f})=
\begin{cases} m_{\rm d}\left(\frac{r_{\rm f}}{r_{\rm d}}\right)^2 & \mbox{ if }r<r_{\rm d} \\ m_{\rm d} & \mbox{ if }r\geq r_{\rm d}
\end{cases}
\end{equation*}
We obtain the $r_{\rm f}$ value associated to each $r_{\rm i}$ solving numerically Equation \ref{eq:AC}, and naturally obtain the adiabatically contracted dark matter profile $M_{\rm dm}(r_{\rm f})$ using Equation \ref{eq:AC1}. Finally, we estimate the dark matter density profile after contraction as 
\begin{equation}
 \rho_{\rm dm}(r)=\frac{1}{4\pi r^2}\frac{dM_{\rm dm}(r)}{dr}.
\end{equation}


\label{lastpage}
\end{document}